\documentclass{PoS}
\usepackage{float}

\newcommand{\be}{\begin{eqnarray}}
\newcommand{\ee}{\end{eqnarray}}
\newcommand{\ba}{\begin{array}}
\newcommand{\ea}{\end{array}}

\newcommand{\bea}{\begin{eqnarray}}
\newcommand{\eea}{\end{eqnarray}}

\newcommand{\bi}{\begin{itemize}}
\newcommand{\ei}{\end{itemize}}

\newcommand{\nn}{\nonumber}

\title{Nucleon-to-meson transition distribution amplitudes in impact parameter space}

\ShortTitle{Nucleon-to-meson TDAs in impact parameter space}

\author{Bernard Pire\\
        CPHT, CNRS, \'Ecole Polytechnique, I. P. Paris, FR-91128 Palaiseau,     France\\
        E-mail: \email{bernard.pire@polytechnique.edu}}

\author{\speaker{Kirill Semenov-Tian-Shansky}
\\
       National Research Centre ``Kurchatov Institute'': Petersburg Nuclear Physics
Institute, RU-188300 Gatchina, Russia\\
        E-mail: \email{cyrstsh@thd.pnpi.spb.ru}}

\author{Lech Szymanowski\\
         National Centre for Nuclear Research (NCBJ), Pasteura 7, PL-02-093 Warsaw, Poland\\
        E-mail: \email{Lech.Szymanowski@ncbj.gov.pl}}

\abstract{
Recent analyses of  backward meson electroproduction support the validity of a collinear QCD factorization framework for these hard exclusive reactions. This opens a way to the extraction of nucleon-to-meson transition distribution amplitudes (TDAs) from the experimental data. Similarly to the generalized parton distributions, TDAs - after the Fourier transform in the transverse plane - carry  valuable information on the transverse location of hadron constituents.
We address the properties of integrated nucleon-to-meson TDAs in the impact parameter representation. We argue that the emerging picture
provides an intuitive interpretation for the hadron structural information contained in
nucleon-to-meson TDAs and allows to study diquark-quark contents of fast moving hadrons in the transverse plane.  }

\FullConference{Light Cone 2019 - QCD on the light cone: from hadrons to heavy ions - LC2019\\
		16-20 September 2019\\
		Ecole Polytechnique, Palaiseau, France}

\begin{document}

\section{Introduction}

Nucleon-to-meson Transition Distribution Amplitudes (TDAs) 
\cite{Frankfurt:1999fp,Pire:2004ie}
occur within the collinear factorized description of a class of hard exclusive reactions with a non-zero baryon number exchange in the cross channel. Prominent examples of such reactions are the backward hard electroproduction of mesons off nucleons 
\cite{Lansberg:2011aa,Pire:2015kxa}
and nucleon-antinucleon annihilation into a lepton pair (or a heavy quarkonium) associated with a meson \cite{Lansberg:2012ha,Pire:2013jva}. According to the usual logic
of the QCD collinear factorization approach, nucleon-to-meson TDAs are universal non-perturbative objects defined as nucleon-meson matrix elements of a three-quark light-cone operator. Nucleon-to-meson TDAs share common features both with baryon Distribution Amplitudes (DAs)  and with Generalized Parton Distributions (GPDs) (see respectively Refs.~\cite{Braun:1999te} 
and
\cite{Diehl:2003ny} 
for reviews).

Recent experimental studies
\cite{Park:2017irz,Li:2019xyp,SD}
brought first evidences in favor of the validity of the reaction mechanism involving nucleon-to-meson TDAs for the description of backward pion and
$\omega$-electroproduction at JLab kinematical conditions.
The perspective to access nucleon-to-meson TDAs experimentally rises a high
demand for refining their physical contents.
In this paper, relying on the similarity between nucleon-to-meson TDAs and GPDs,  
we build an intuitive physical picture for TDAs
in the impact parameter space.

\section{Integrated $\pi N$ TDAs and quark-diquark picture of the nucleon}

For definiteness, throughout this paper, we consider the case of the
proton ($N^p$)-to-$\pi^0$ $uud$ TDAs. However, our results admit a straightforward
generalization for other isospin channels for
$\pi N$
TDAs as well as for more involved cases. The leading twist-$3$ proton-to-$\pi^0$ TDAs are defined through the Fourier transform of $N^p$-$\pi^0$ matrix element of the $uud$ trilocal operator on the light cone ($n^2=0$)%
\footnote{The explicit form of the parametrization of the 
leading twist-$3$ $\pi N$ TDAs is given in Eq.~(10) of Ref.~\cite{Pire:2011xv}.
Here, for simplicity, we assume that the overall normalization factor $i\frac{f_N}{f_\pi M_N}$
is included into the definition of invariant TDAs.}:
\be
&&
4 
(P \cdot n)^3 \int \left[ \prod_{i=1}^3 \frac{d \lambda_i}{2 \pi}  \right]
e^{i(\lambda_1 x_1+\lambda_2 x_2+\lambda_3 x_3) (P \cdot n)}
\langle \pi^0(p_\pi) | \, u_\rho(\lambda_1 n) u_\tau(\lambda_2 n) d_\chi(\lambda_3 n) \,
|N^p(p_N;\,s_N) \rangle
\nn \\ &&
=  \delta(x_1+x_2+x_3-2\xi) 
\sum_{H= V_{1,2},\,A_{1,2},\atop T_{1,2,3,4}}
h_{\rho \tau, \, \chi}^H
H^{\pi^0 N^p}(x_1,x_2,x_3,\xi,\Delta^2).
\label{Master_Fourier}
\ee
Here
$n$
is the light-cone vector
$n^2=0$.
For simplicity we adopt the light-like gauge
$A \cdot n=0$
and therefore omit the gauge links in the trilocal quark operator. Antisymmetrization in quark color
indices (which we do not show explicitly) is assumed. The notations for
kinematical variables follow the usual conventions:
$P= \frac{p_N+p_\pi}{2}$
is the average momentum,
$\Delta=p_\pi-p_N$
is the momentum transfer between the meson and the nucleon; $s_N$ denotes
the nucleon polarization variable. 
The sum in the r.h.s. of (\ref{Master_Fourier})
stands over the $8$ independent Dirac structures
$h_{\rho \tau, \, \chi}^H$
relevant at the leading twist-$3$ accuracy.
Each of the $8$ proton-to-$\pi^0$ leading twist-$3$
$uud$
TDAs are functions of the
$3$ light-cone momentum fractions
$x_i$,
the skewness variable
$\xi= - \frac{\Delta \cdot n}{2 P \cdot n}$
defined with respect to the longitudinal momentum transfer, momentum transfer squared $\Delta^2$
as well as of the factorization scale. 

\begin{figure}[H]
 \begin{center}
 \includegraphics[width=.20\textwidth]{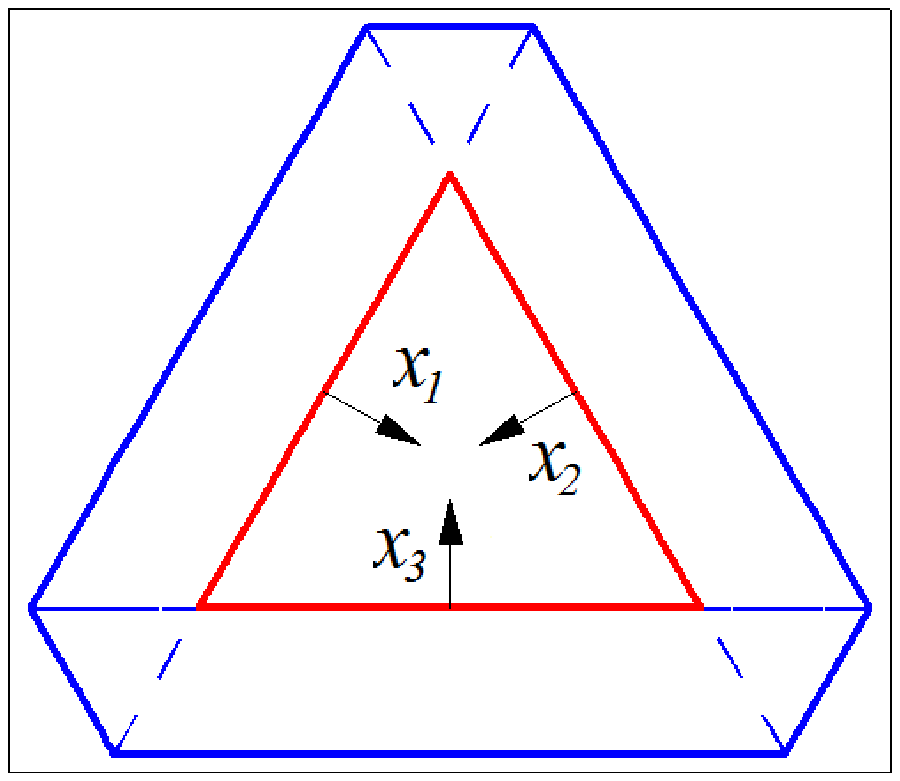}
 \includegraphics[width=.20\textwidth]{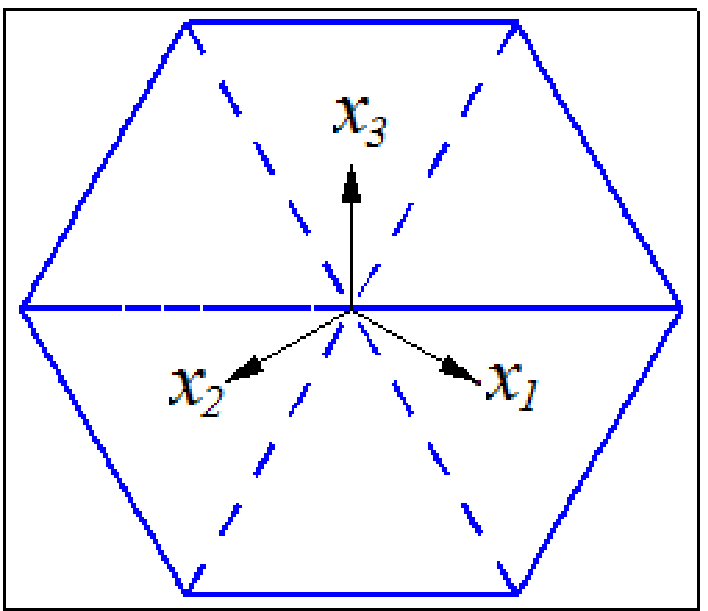}
 \includegraphics[width=.20\textwidth]{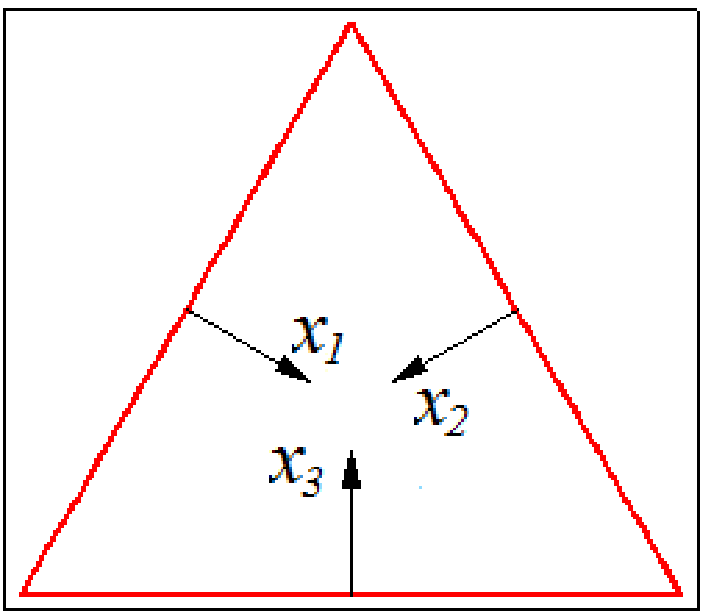}
  \end{center}
\caption{Physical domains for $\pi N$ TDAs in the
barycentric coordinates $\sum_i x_i=2\xi$. {\bf Left panel:} general case  $0<\xi <1$.
{\bf Central panel:} case $\xi=0$. {\bf Right panel:} case $\xi=1$.}
\label{Fig_Domain}
\end{figure}

The support domain of nucleon-to-meson TDAs in the momentum fraction variables $x_i$ has been worked out in Ref.~\cite{Pire:2010if}. In the barycentric coordinates defined by the
longitudinal momentum constraint 
$\sum_i x_i=2 \xi$ 
it is given by the intersection of
$3$ stripes 
$-1+\xi \le x_i \le 1+\xi$. Similarly to the GPD case, it is natural to single out the Efremov-Radyushkin-Brodsky-Lepage (ERBL)-like region, in which all three quark longitudinal momentum fractions are positive, and
two types of Dokshitzer-Gribov-Lipatov-Altarelli-
Parisi (DGLAP)-like regions, where either one or two quark longitudinal momentum fractions
are positive and the remaining two or one are negative (see Fig.~\ref{Fig_Domain}). The DGLAP-like-I,II
and ERBL-like domains are separated by the cross-over lines $x_i=0$. Note, that, contrary to the GPD case, the complete support of nucleon-to-meson TDAs depends on $\xi$ (see Fig.~\ref{Fig_Domain}).

It turns out to be convenient to switch to two independent momentum
fraction variables instead of $x_i$ that are subject to the constraint $\sum_i x_i=2 \xi$.
A natural choice of independent variables is given by the so-called quark-diquark coordinates $(w_i, \, v_i)$ (there exist $3$ equivalent choices $i=1,\,2,\,3$ of quark-diquark coordinates, depending on which pair of quark momenta is selected to constitute the momentum
of a diquark):
$
w_i=x_i-\xi; \ \  v_i= \frac{1}{2} \sum_{k,l=1}^3 \varepsilon_{ikl} x_k.
$
Within these coordinates the support of nucleon-to-meson TDAs can be parameterized as
\be
-1 \le w_i \le 1; \ \ \ -1+|\xi-\xi'_i| \le v_i \le 1- |\xi-\xi'_i| \ \
{\rm with} \ \ {\xi'}_i \equiv  \frac{\xi -w_i}{2}.
\label{Support_quark_diquark}
\ee

In order to make contact to a quark-diquark picture we
would now like to introduce $\pi N$ TDAs integrated over
the momentum fraction variable $v_i$.
For definiteness we choose it to be  $v_3 \equiv \frac{x_1-x_2}{2}$.
We employ
the following identity for the exponent in (\ref{Master_Fourier}):
\be
e^{i(x_1 \lambda_1+ x_2 \lambda_2+ x_3 \lambda_3) P \cdot n}=
e^{i( \frac{x_1-x_2}{2} (\lambda_1-\lambda_2) +
\frac{x_1+x_2}{2} (\lambda_1+\lambda_2)
+ x_3 \lambda_3) P \cdot n} \equiv
e^{i( v_3 (\lambda_1-\lambda_2) +
\xi_3' (\lambda_1+\lambda_2)
+ x_3 \lambda_3) P \cdot n}.
\ee
Then we integrate the equality (\ref{Master_Fourier})  in $v_3$ over
the interval $(-\infty; \, \infty)$.
In the r.h.s. we may integrate just over the $v_3$ support
(\ref{Support_quark_diquark})
of  $\pi N$ TDAs. In the l.h.s. the integration of the exponent results in the delta function $\delta(\lambda_1- \lambda_2)$.
This latter delta function can be employed to
remove one of the
$\lambda$-integrals. It brings the arguments of the two
$u$-quark operators to the
same point on the light cone
$\lambda_1 n = \lambda_2 n \equiv \lambda_D n$
and gives rise to the hadronic matrix element of a bilocal light-cone operator:
\be
\widehat{O}_{\rho \tau \chi}^{\{uu\}d}( \lambda_D n, \lambda_D n,\lambda_3 n)=
u_\rho(\lambda_D n) u_\tau(\lambda_D n) d_\chi(\lambda_3 n) \equiv 
\widehat{D}_{\rho \tau }( \lambda_D n) d_\chi(\lambda_3 n).
\label{diquark_quark_op_uud}
\ee
It is natural to interpret this operator as the bilocal $uu$-diquark- $d$-quark 
operator on the light-cone. Now we employ the translation invariance of the
matrix element to translate the arguments of the bilocal operator 
(\ref{diquark_quark_op_uud}) 
to the symmetric points 
$\pm \frac{\lambda}{2}$ on the light-cone introducing $\lambda \equiv \lambda_3-\lambda_D$ and $\mu \equiv \lambda_3+\lambda_D$. The integral
over $\mu$ can be performed producing the momentum conservation $\delta$-function:
\be
\int \frac{d \mu}{2 \pi}
 e^{i  \left( x_D+x_3 -2\xi  \right) (P \cdot n) 2 \mu}= \frac{1}{2(P \cdot n)}
 \delta(x_D+x_3-2\xi); \ \ \ x_D \equiv x_1+x_2=2 \xi'_3.
\ee
Finally, we get the equality relating the Fourier transform of the nucleon-pion matrix element
of the light-cone diquark-quark operator
(\ref{diquark_quark_op_uud})
to the $v_3$-integrated nucleon-to-pion TDAs
\be
&&
2 (P \cdot n)
\int  \frac{d \lambda}{4 \pi}
 e^{i  \left( w_3 \lambda  \right) (P \cdot n) }
 \langle \pi(p_\pi)|
  \widehat{D}_{\rho \tau}^{\,uu}(- \frac{\lambda}{2} n)
 {d}_\chi ( \frac{\lambda}{2} n)| N(p_N) \rangle \nn
\\ &&
=
%
\sum_{H=V_{1,2}, \, A_{1,2}, \, T_{1,2,3,4}}  h^H_{\rho \tau, \, \chi}
\int_{-1+|\xi-\xi'_3|}^{1-|\xi-\xi'_3|} dv_3 
H^{\pi^0 N^p}(w_3, \, v_3, \, \xi, \, \Delta^2) \,.
\label{TDA_result_integrated}
\ee
The $v_3$-integrated TDAs occurring in the r.h.s. of   
Eq.~(\ref{TDA_result_integrated})
share many common features with GPDs. Namely, they are functions of
one longitudinal momentum fraction $w_3= \frac{x_3-x_D}{2} \in [-1;\, 1]$,
of skewness $\xi$, of invariant momentum transfer $\Delta^2$ and of factorization scale.
As a consequence of the Lorentz invariance the Mellin moments
of $v_3$-integrated TDAs possess the usual polynomiality property in $\xi$. 
For the $v_3$-integrated TDAs it is natural to specify the ERBL-region with $w_3 \in [-\xi;\xi]$,
DGLAP-I region $w_3 \in [-1;\,-\xi]$ and DGLAP-II region $w_3 \in [\xi;\, 1]$.
We now propose an interpretation of these objects in the impact
parameter space. This allows to use $v$-integrated TDAs as a tool to study 
the quark-diquark structure of hadrons in the transverse plane.
It is worth emphasizing that, contrary to GPDs, the $v$-integrated TDAs do not
possess a comprehensible forward limit in which a probabilistic interpretation 
\cite{Burkardt:2000za,Burkardt:2002hr} 
applies for GPDs. Thus, similarly to the case of GPDs
with non-zero skewness, we get an interpretation in terms of 
the probability amplitudes.

\section{Integrated $\pi N$ TDAs in impact parameter space}
The use of the impact parameter representation for building up
a vivid physical picture of hadrons in the transverse
plane has been pioneered by M.~Burkardt for GPDs in the zero
skewness limit in Refs.~\cite{Burkardt:2000za,Burkardt:2002hr}.
The extension of this framework for GPDs with non-zero skewness
was proposed in Refs.~\cite{Ralston:2001xs,Diehl:2002he}.
Since the general structure of $v$-integrated nucleon-to-meson TDAs looks similar to
GPDs it is natural to adopt for them the impact parameter space representation.

The first step consists in introducing the initial nucleon and final meson states
with specified longitudinal momenta localized around a definite position $\bf b$
in the transverse plane:
\be
&&
|p_N^+,\, {\bf b}; \, s_N \rangle=  \int \frac{d^2 {\bf p}_N}{16 \pi^3} e^{-i {\bf p}_N \cdot {\bf b}} |p_N^+,\, {\bf p}_N; \, s_N \rangle; \ \ \ 
\langle p_\pi^+,\, {\bf b} |= \int \frac{d^2 {\bf p}_\pi}{16 \pi^3}
e^{i {\bf p}_\pi \cdot {\bf b}}
\langle p_\pi^+,\, {\bf p}_\pi |.
\label{Loc_states}
\ee
Rigorous treatment requires forming wave packets with precisely localized states
(\ref{Loc_states})
using a smooth weight falling sufficiently fast at infinity with $|{\bf p}|$
in order to avoid infinities due to normalization. A possible choice is to
employ a Gaussian wave packets with the same standard deviation
parameter for the initial nucleon and the final meson.  

Switching to the impact parameter space representation is performed by
Fourier transforming the corresponding operator hadronic matrix element with respect
to the transverse component
$\bf D$
of the vector
$
D=\frac{p_\pi}{1-\xi}- \frac{p_N}{1+\xi}
$.
By construction, the transverse component
$\bf D$
is invariant under the transverse boosts. This ensures
that the invariant momentum transfer  depends on
${\bf p}_N$
and
${\bf p}_\pi$
only through $\bf D$:
$
\Delta^2=-2\xi \left( \frac{m_\pi^2}{1-\xi} - \frac{M_N^2}{1+\xi} \right)- (1-\xi^2) {\bf D}^2.
$

We consider the matrix elements of the diquark-quark
operator with the explicit dependence on the transverse position $\bf{z}$:
\be
{\widehat {\cal O}}_{\rho \tau \chi}^{\{uu\} d}({\bf{z}})= \int \frac{d \lambda}{4 \pi}
e^{i  \left( w_3 \lambda  \right) (P \cdot n) }
 {u}_{\rho}(0,\,- \frac{\lambda}{2}, \, {\bf z})
 {u}_{\tau}(0,\,- \frac{\lambda}{2}, \, {\bf z})
 {d}_\chi (0,\, \frac{\lambda}{2} n, \, {\bf z}),
 \label{Dq_operator_trans}
\ee
where we adopt the following convention for the
position arguments of quark fields: $q(z)=q(z^+,\,z^-,\,{\bf z})$.

In order to single out a particular combination of $\pi N$ TDAs
we contract the matrix element of the operator
(\ref{Dq_operator_trans})
over the Dirac indices with a suitable projector.
Assuming that for $\pi N$ TDAs we employ the parametrization of Eq.~(10) of Ref.~\cite{Pire:2011xv}
we, as an example, have chosen to contract the matrix element (\ref{Dq_operator_trans}) with%
\footnote{We employ Dirac's ``hat'' notation $\hat{a} \equiv \gamma_\mu a^\mu$; $C$ is the charge conjugation matrix.}
$
v^{-1}_{\tau \rho, \, \chi}=\left( C^{-1} \hat{P} \right)_{\tau \rho} \bar{U}_\chi(p_N,s_N).
$
Since we deal with unpolarized nucleon, we sum and average over the nucleon spin $s_N$.
This convolution singles out the following combination of $v_3$-integrated $\pi N$ TDAs:
\be
&&
{\cal H}^{\pi N}(w_3, \,\xi,\, {\bf D})=
 (4 M_N^2-\Delta^2) 
\int_{-1+|\xi-\xi'_3|}^{1-|\xi-\xi'_3|} dv_3
\frac{1}{2} \Big[ (3M_N^2+m_\pi^2-\Delta^2) V_1^{\pi N}(w_3,\,v_3,\,\xi,\, \Delta^2)
\nn \\ &&
+ (-2\Delta^2-2M_N^2+2m_\pi^2) V_2^{\pi N}(w_3,\,v_3,\,\xi,\, \Delta^2)
\Big], 
\label{Comb_H}
\ee 
where $\Delta^2$ is expressed through ${\bf D}^2$. 
The contraction with different projectors%
\footnote{A proper design of the projecting operation establishing connection
with corresponding diquark-quark helicity amplitudes still has to be developed.}
 will allow to single out
other combinations of $8$ leading twist-$3$ proton-to-$\pi^0$ $uud$ TDAs.
The transition to the impact parameter space then gives

\be
&&
\int \frac{d^2 { \bf D}}{(2\pi)^2}\; e^{-i\,  {\bf D}\cdot {\bf b}}\,
      {\cal H}^{\pi N}(w,\xi, {\bf D})
\nonumber \\ &&
= {\cal N}^{-1}\frac{1+\xi^2}{(1-\xi^2)^{2}}\:\; \sum_{s_N}
\left( C^{-1} \hat{P}  \right)_{\tau \rho}  \bar{U}_\chi(p_N; \, s_N)
   \Big\langle p_\pi^+,
     -\frac{\xi  {\bf b}}{1-\xi} \,
        \Big|\, 
        {\widehat {\cal O}}_{\rho \tau \chi}^{\{uu\} d}({\bf{b}})
        \,\Big|\,
         p^+_N, \frac{\xi  {\bf b}}{1+\xi}; \, s_N \Big\rangle  \;,
  \label{matrix-final_our}
\ee
where $\cal N$ is the normalization factor originating from the normalization
of the localized states (\ref{Loc_states}). Without use of the smooth wave packets  it turns to be singular as $\delta^{(2)}({\bf 0})$.

\begin{figure}[H]
 \begin{center}
 \includegraphics[width=.42\textwidth]{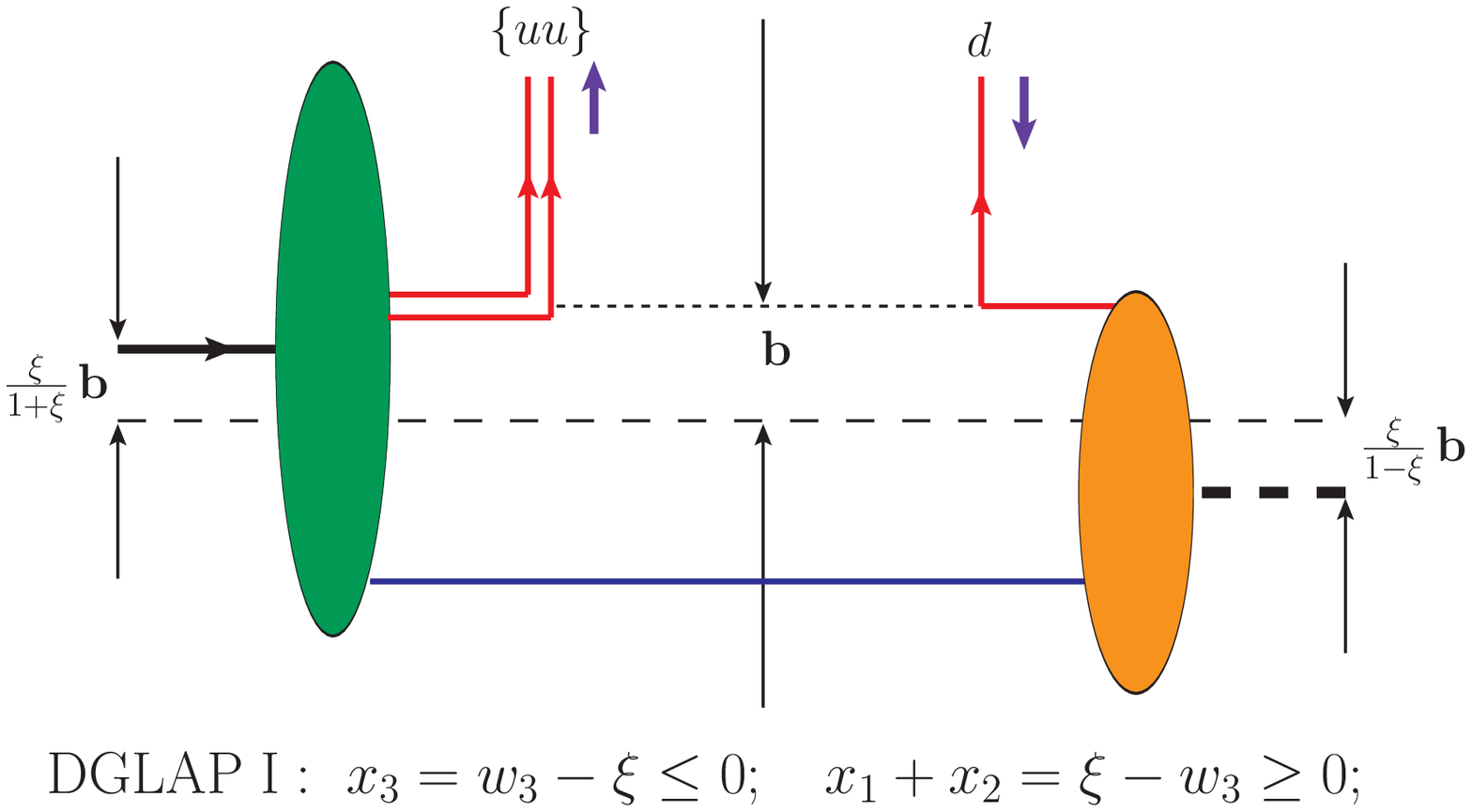}
  \includegraphics[width=.42\textwidth]{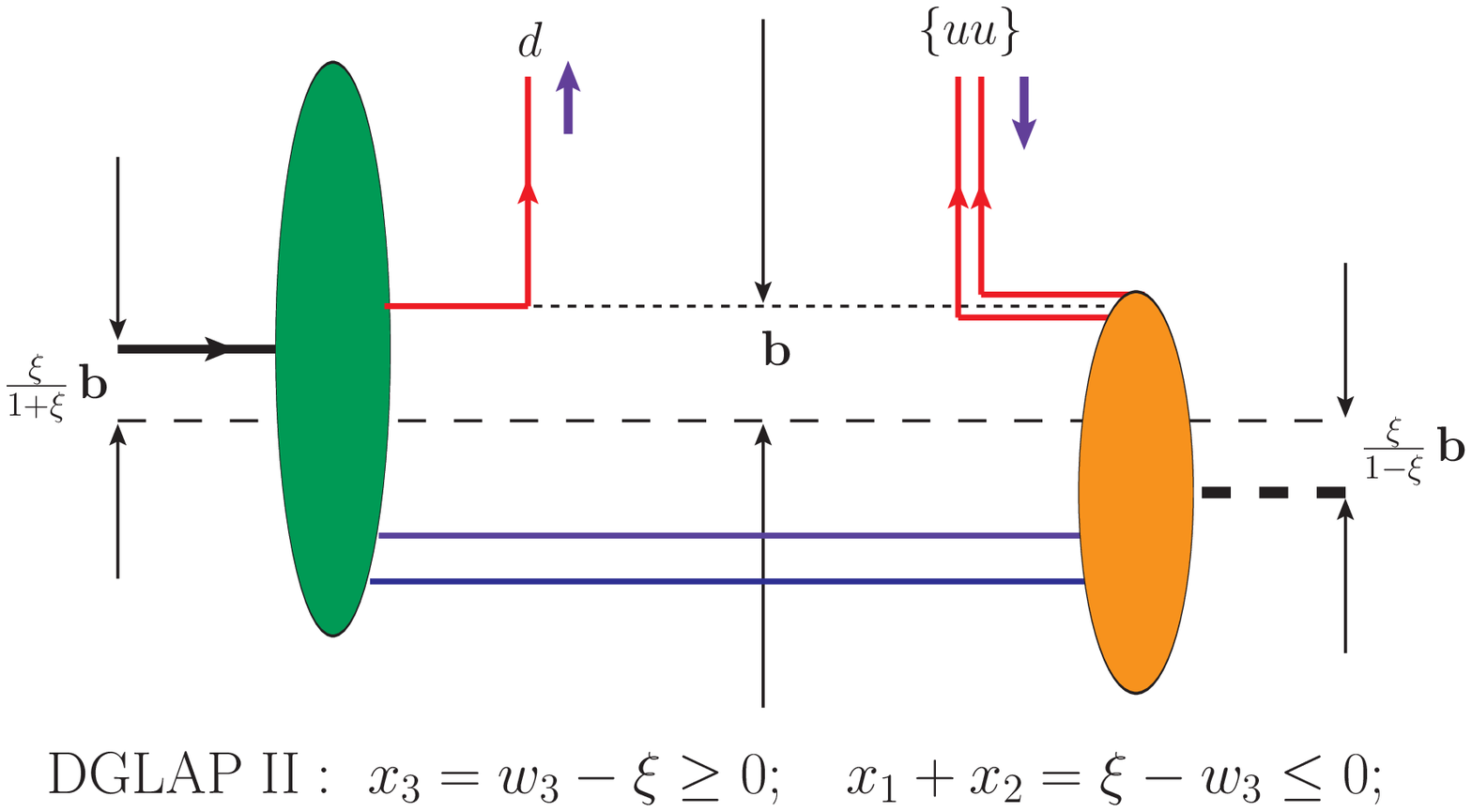} \\
  \vspace{1em}
\includegraphics[width=.42\textwidth]{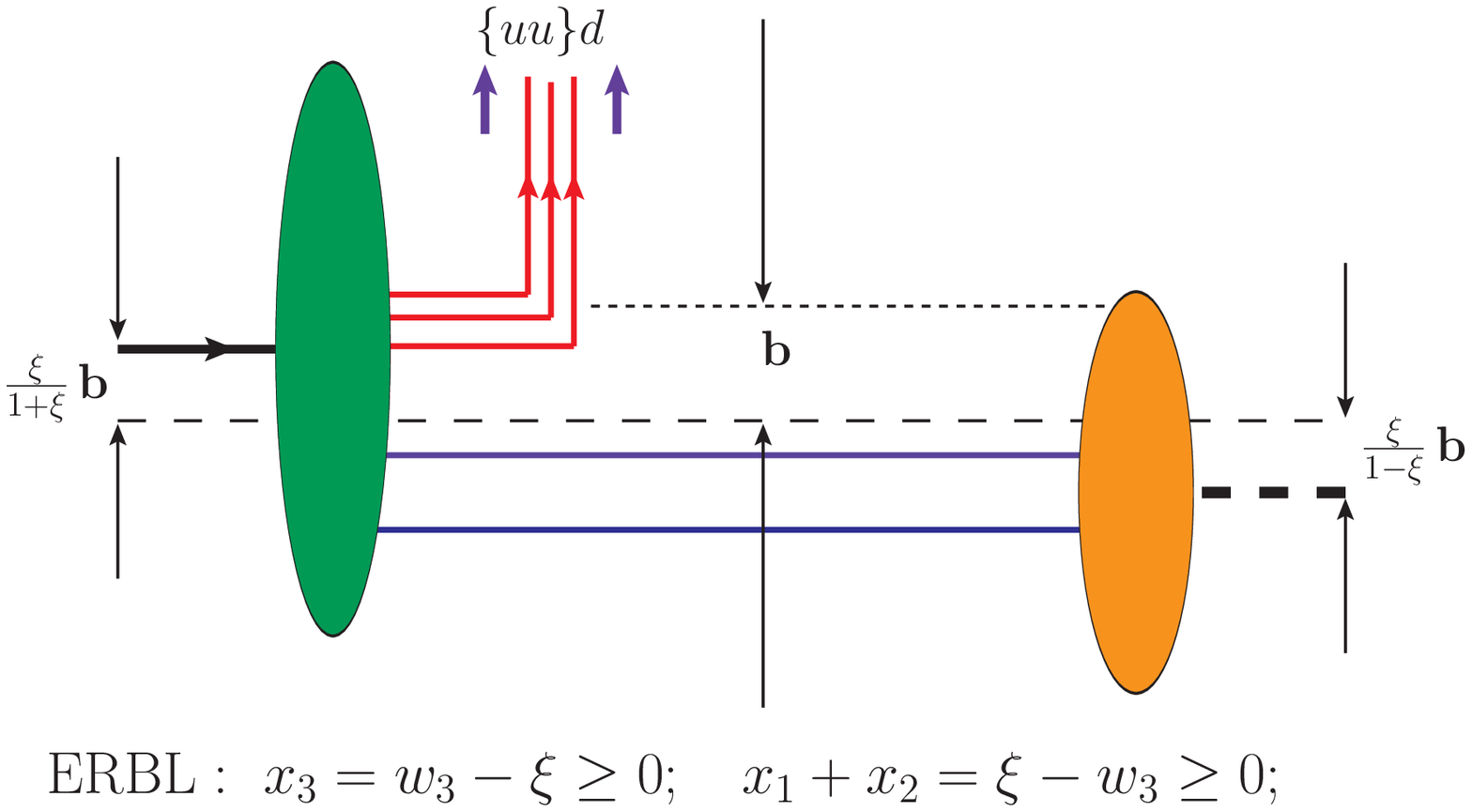}
\end{center}
\caption{Impact parameter space interpretation for the $v_3$-integrated $uud$ $\pi N$ TDA in the DGLAP-like~I, DGLAP-like~II  and in the ERBL-like domains. Solid arrows show the direction of the positive longitudinal momentum flow. }
\label{Fig_DGLAP_I_and_II}
\end{figure}

Thus we end up with a picture that is completely analogous to the GPD case:
the hard probe interacts with a partonic configuration at the transverse
position $\bf b$. The initial state nucleon and the finite state meson are
localized around $\bf 0$, but they are shifted one from another by a transverse
separation of the order $\xi {\bf b}$. This interpretation is presented in 
Fig.~\ref{Fig_DGLAP_I_and_II}. It is also qualitatively
consistent with the low Fock component picture proposed in 
\cite{Pasquini:2009ki}.

\bi
\item In the DGLAP-like~I region $w_3 \le -\xi$ the impact parameter
specifies the location where a $uu$-diquark is pulled out of a proton and
then replaced by an antiquark $\bar{d}$ to form the final state meson.
\item In the DGLAP-like~II region $w_3 \ge \xi$ the impact parameter
specifies the location where a quark $d$ is pulled out of a proton and
then replaced by an antidiquark $\bar{u}\bar{u}$ to form the final state meson.
\item In the ERBL-like region $-\xi \le w_3 \le \xi$ the impact parameter
specifies the location where a three-quark cluster composed of a $uu$-diquark
and a $d$-quark is pulled out of the initial nucleon to form the final state meson.
\ei

A complementary picture can be obtained from the $v_1$-integrated $\pi N$ TDAs.
This corresponds to a diquark constructed out of the third and second quarks ($du$).
It makes sense to perform the Fierz transform (see App.~B3 of \cite{Pire:2011xv}) to the relevant set of the Dirac structures:
$h^{\pi N}_{\rho \tau, \, \chi} \to h^{\pi N}_{\chi \tau, \, \rho}$. The projection $v^{-1}_{\tau \rho, \, \chi}$
then involves a different combination of TDAs. The third possible picture resulting from the $v_2$-integrated $\pi N$ TDAs
should be analogous to the $v_1$-integrated case since it also corresponds to a
$\{ud \}u$-diquark-quark operator. 

\section{Conclusions}
In this paper we propose an interpretation for $v$-integrated nucleon-to-pion TDAs
in the impact parameter space. It offers an intuitive interpretation of the information 
contained in nucleon-to-meson TDAs in terms of the diquark-quark contents of the corresponding hadrons.

This project has received funding from the European Union's Horizon 2020 research and innovation programme under grant agreement No 824093. K.S. was supported by the RSF grant  16-12-10267.
L.S. acknowledges the support by the grant 2017/26/M/ST2/01074 of the National Science Center in Poland. He also thanks the LABEX P2IO the GDR QCD and the French-Polish Collaboration Agreement POLONIUM for support.

\end{document}